\documentclass[aps,prl]{revtex4}
\usepackage{times}
\usepackage{graphicx}

\begin{document}

\title{Study of the Villin Headpiece folding dynamics by combining
coarse-grained Monte Carlo evolution and all-atom Molecular Dynamics}

\author{Giacomo M. S. De Mori$^1$ Giorgio Colombo$^1$ and Cristian Micheletti$^2$\\
\small $^1$ Istituto di Chimica del Riconoscimento Molecolare,
CNR, Via Mario Bianco 9, 20131 Milano, Italy\\
\small $^2$ International School for Advanced Studies (S.I.S.S.A.) and INFM, Via
Beirut 2-4, 34014 Trieste, Italy}

\date{\today}

\begin{abstract}
The folding mechanism of the Villin headpiece (HP36) is studied by
means of a novel approach which entails an initial coarse-grained
Monte Carlo (MC) scheme followed by all-atom molecular dynamics (MD)
simulations in explicit solvent. The MC evolution occurs in a
simplified free-energy landscape and allows an efficient selection of
marginally-compact structures which are taken as viable initial
conformations for the MD.  The coarse-grained MC structural
representation is connected to the one with atomic resolution through
a ``fine--graining'' reconstruction algorithm. This two-stage strategy
is used to select and follow the dynamics of seven different unrelated
conformations of HP36. In a notable case the MD trajectory rapidly
evolves towards the folded state, yielding a typical RMS deviation of 
the core region of only 2.4\AA\ from the closest NMR model (the 
typical RMSD over the whole structure being 4.0\AA). The analysis of
the various MC-MD trajectories provides valuable insight into the
details of the folding and mis-folding mechanisms and particularly
about the delicate influence of local and non-local interactions in
steering the folding process.
\end{abstract}

\maketitle

\section*{Introduction}
The characterization of the process by which proteins fold into their
native structures is at the heart of molecular biology. Starting from
the pioneering work of Anfinsen, several experimental studies in the
past decades have provided support for the fact that the knowledge of
the mere primary sequence ought to allow the prediction of the native
structure for a large class of proteins
\cite{anfinsen,anfinsenscheraga,tooze,creighton,fersht}. This
conclusion relies on the notion that, under physiological conditions,
the native conformation is at a minimum of a multidimensional
free-energy landscape
\cite{anfinsen,anfinsenscheraga,Funnel1,Funnel2,Funnel3,Funnel4,Funnel5}.
Furthermore, the fact that several proteins are known to reach
the native configuration from a variety of initial configurations has
been rationalized in the assumption that, through natural evolution,
the free-energy profile around the native minimum is relatively smooth
and with a wide basin of attraction in configuration space
\cite{Levinthal,Funnel3,Miche99b,Mirny98}.

These facts open, in principle, the possibility to predict the native
state of proteins through a numerical simulation of the dynamical
evolution of an arbitrary initial (``unfolded'') conformation. The
feasibility of this scheme would have profound ramifications both from
the theoretical and practical point of view. Though there is an
ongoing progress in this direction
\cite{CASP1to4,CASP5a,CASP5b,Scheraga_protA}, present computational
techniques and resources are not adequate to follow the folding
dynamics of a medium-size protein (e.g. 100 residues) in its
surrounding solvent for time intervals comparable to the experimental
folding
times\cite{Berendsen_grail,Brooks_fold_unfold,Dobson98,Karplus02}.

The methodology that we discuss here aims at augmenting the reach of
numerical folding dynamics starting from the only knowledge of the
protein sequence. The strategy exploits a two-stage simulation scheme
where a protein's dynamical evolution is first followed through a
coarse-grained Monte Carlo (MC) conformational search followed by all
atom molecular dynamics (MD) simulations in explicit solvent. We
discuss its application to the Villin headpiece (HP36) for which we
generate several independent trajectories (and reach, for the core
region, residues 9--32, a root mean square deviation (RMSD) of 0.24 nm
from the closest NMR model\cite{McKnightNSB}).

The interest in the villin headpiece stems from the fact that, despite
it consists of only 36 residues, with no cysteines, it is able to fold
autonomously and rapidly (in about 10 $\mu$s
\cite{EatonJMBVillin,McKnightVillin}).  Along with other short
fast-folding proteins, it has been extensively investigated
experimentally with the purpose of elucidating general mechanisms of
the folding kinetics that ought to be transferable to longer and more
complex proteins. On the other hand villin also
constitutes an ideal reference case for theoretical studies, since its
short length and folding rapidity allow to bridge the gap between the
time scales addressable by explicit numerical dynamics and those
relevant for
folding\cite{Duan98,BakerKollmanJMB,SpoelVillin,BakerKollmanJACS,Hansmann2Villin,Freed02,ScheragaVill04}.

In this respect, a notable reference study is constituted by the one
of Duan and Kollman which, starting from an extended villin
conformation, followed its evolution for about 1 $\mu$s using the
AMBER force field and explicit solvent\cite{Duan98}. It is important
to remark that the explicit modeling of water molecules in the
simulation introduces a number of atoms that typically greatly exceeds
the ones constituting the protein itself and hence is computationally
very intensive. This computational cost is rewarded by the possibility
to capture the delicate, but crucial effects of the solvent-protein
interactions\cite{Wolynesacqua}.

By relaxing the accuracy of the solvent treatment it is possible to
increase the MD efficiency. This approach, which usually entails the
``integration'' of the solvent degrees of freedom in effective
inter-atomic potentials has been adopted in several simulation
studies. Ferrara and Caflisch, for instance, could follow the folding
pathway of a three-stranded antiparallel ${\beta}$-sheet peptide at
high temperature with MD \cite{ferrcaf01}. More recently Pande and
Shirts adopted the implicit solvent approximation in an innovative
distributed-computing approach \cite{Shirts}. For the case of Villin,
the numerical speed-up obtained by the simplified solvent treatment
allowed to address a total time scale of the order of 1 $\mu$s,
distributed across a large number of short trajectories each of about
30 ns. In such short simulations the choice of the starting protein
structures has a profound impact on the possibility to relate the
properties of the simulated system to experiments. This issue has been
recently addressed by Fersht who discussed the presence of important
lag--phases in all--atom MD simulations starting from random or
completely extended conformations \cite{Fershtdistributed,Paci2003}.

Several methods and techniques have been developed over the years to
overcome the time limitations of all-atom MD. A widely-used strategy
is to renounce to follow the folding pathway in favor of identifying
the conformations of lowest free energy through various optimization
techniques including simulated annealing, multicanonical methods,
genetic algorithms etc.
\cite{moultgenetic,TROW,confspaceanneal,irback95,Hansmann99,entropysampl,mucapept,garciaonuchic03}.
Another commonly employed route is to simplify not only the
description of the solvent but also of the protein structure itself.
Accordingly, several coarse-grained models have been introduced where
the protein is described as a chain of linked beads which interact
through suitable effective potentials. The drastic structural
simplification entailed by these models allows to explore a portion of
the conformational space vastly larger than in all-atom MD. However,
the limitations of both the structural representations and of the
energy functions clearly prevent from reproducing the finer features
of the folding process. Nevertheless they have proved valuable for
capturing and analyzing various general aspects such as the
calorimetric cooperativity, the relation between native-state topology
and folding rates, the identification of folding nuclei and the
modeling of functional motions etc.
\cite{bakernature,Miche99b,Clem,thirum,cieplak,scheraga,Kolinskyskolnick94b,jchemphys2,karp,hammes44,hoang2000,hoang2000b,irback02}

These observations suggested the present approach which combines a
coarse-grained Monte Carlo (MC) search with all atom molecular
dynamics (MD) simulations in explicit solvent. The effect of the
coarse-grained part is to simplify the protein's energy landscape so
to identify efficiently physically-meaningful starting conformations
for the MD.  After bringing these structures into the realm of
all-atom representations, explicit solvent MD is used to introduce
back the fine chemical details which are ultimately responsible for
driving the evolution towards the native state. The link between the
two structural representations is a fine-graining algorithm which
allows to reconstruct reliably the full atomic detail of the protein
using a library of previously generated protein fragments. The scheme
adopted here therefore aims at extending the reach of ordinary MD
simulations by using simple physico-chemical criteria to identify the
marginally-compact starting conformations in which partial formation
of secondary and tertiary interactions has taken place. Thus, the
approach has a different spirit from that followed in other two-stage
approaches, notably the pioneering work of Vieth et al. \cite{vieth2stage}
and Liwo et al. \cite{liwo2stage}, where the fine-graining step was
aimed at perturbing or refining the coarse-grained structures which
minimized a given energy functional

\section*{Results}

\begin{figure}
\includegraphics[width=0.7\textwidth]{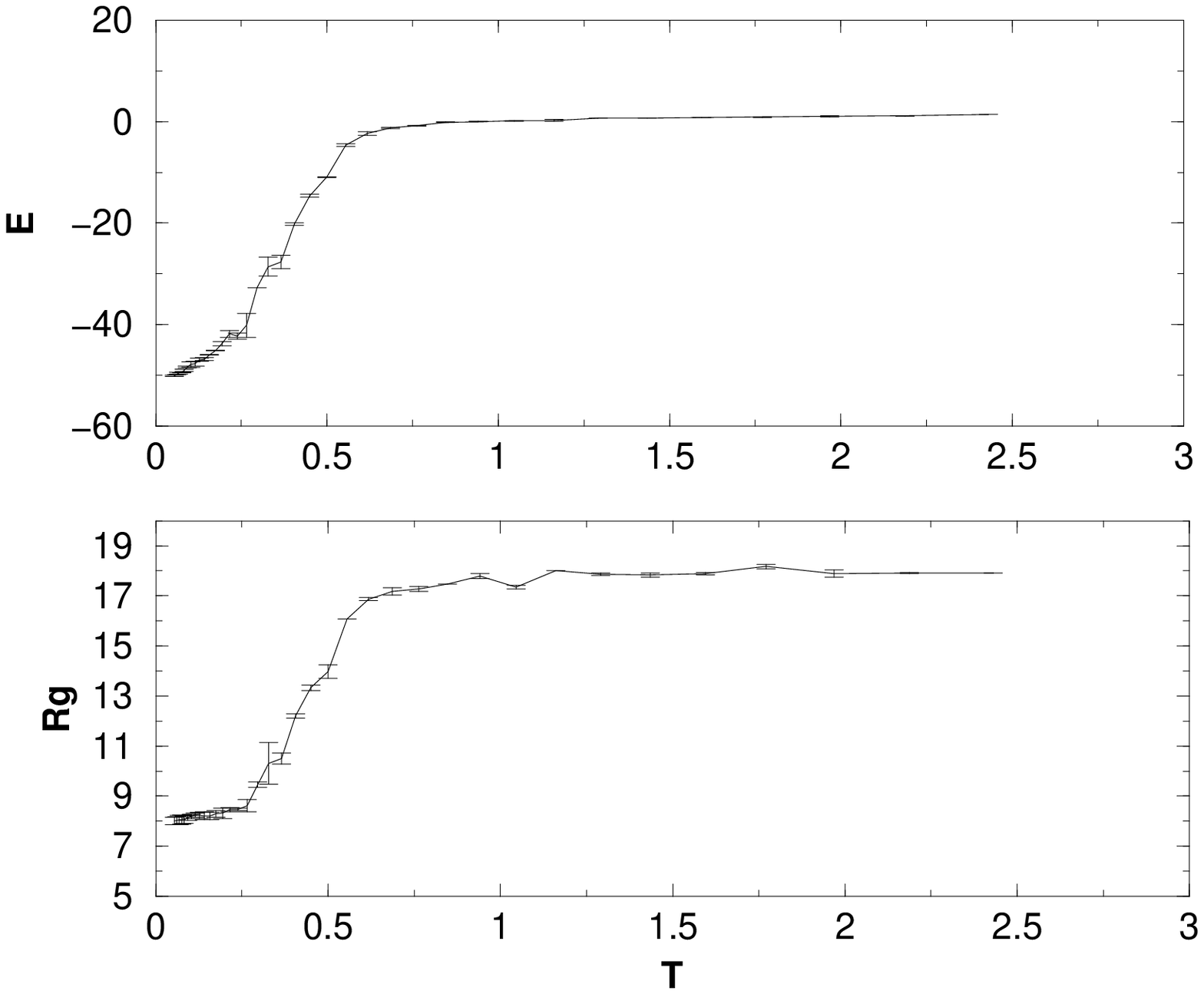}
\caption{Average energy and radius of gyration as a function of
temperature for the coarse-grained model. For each temperature the
averages were calculated over 1000 uncorrelated structures.}
\label{fig:mc}
\end{figure}

The simplification of the free-energy landscape operated by the
coarse-grained MC approach is achieved through a coarse-graining of
the structural degrees of freedom of the protein; in particular the
protein is described in terms of its $C_\alpha$ trace and of the
associated $C_\beta$ centroids. This schematic description is
accompanied by a simplification of the energy functional (see Methods)
which incorporates effective pairwise interactions among amino acids,
the local propensity to form secondary motifs and a term favoring the
packing of the latter. Within this framework, the thermodynamics of
the HP36 protein can be characterized by means of a MC evolution
involving crankshaft and pivot moves. These moves preserve the length
of the effective bonds joining consecutive $C_\alpha$'s (equal to
3.8\AA). The average energy and radius of gyration, $R_g$ as a
function of temperature of the coarse-grained system are visible in
Fig. \ref{fig:mc}.

As the temperature of the system is decreased, the protein undergoes a
collapse signaled by the concomitant decrease of both the radius of
gyration, $R_g$, and the average system energy, see Fig. \ref{fig:mc}.
The system density of states does not however possess the marked
concavity which is distinctive of first-order transitions. This fact
reflects the generality of the coarse-grained energy functional
adopted here. In fact, in order to reproduce in simplified models the
typical all-or-none character of folding transitions it is necessary
to optimize either the protein sequence or the energy parameters
\cite{scheraga,kaya1,kaya2,Hao98,shak_gutin_93}. In correspondence of
the collapse temperature, $T_c$, the specific heat exhibits a peak
which reflects the large energy fluctuations associated to the
coexistence of rather open configurations with more globular
structures (see Fig. \ref{fig:mc}). The latter ones typically possess
local secondary elements (up to 1/3 of the residues are found in
helical conformations) and are further compactified at lower
temperatures. The protein at $T_c$ is thus poised to collapse into
compact conformations with non-trivial secondary content. Thus, the
structures sampled by the MC algorithm at $T_c$ represent attractive
candidates for all-atom MD evolution for several reasons: (1) they
possess good secondary content, (2) they are not unnaturally compact
and (3) the conformational variability at $T_c$ is such that
significantly different structures can be picked.  It is also worth to
mention that another intuitive strategy would be to start the MD runs
from the lowest energy states of the coarse-grained model. However,
the resulting structures, once reconstructed with their atomic detail,
produced steric clashes among the atoms which could not be removed by
e.g. relaxing the reconstructed structure before the actual MD
run. Accordingly, the lowest-energy conformations of the
coarse-grained model were not used as inputs for the all-atom
dynamical evolution.

\begin{figure}[h!]
\includegraphics[width=0.7\textwidth]{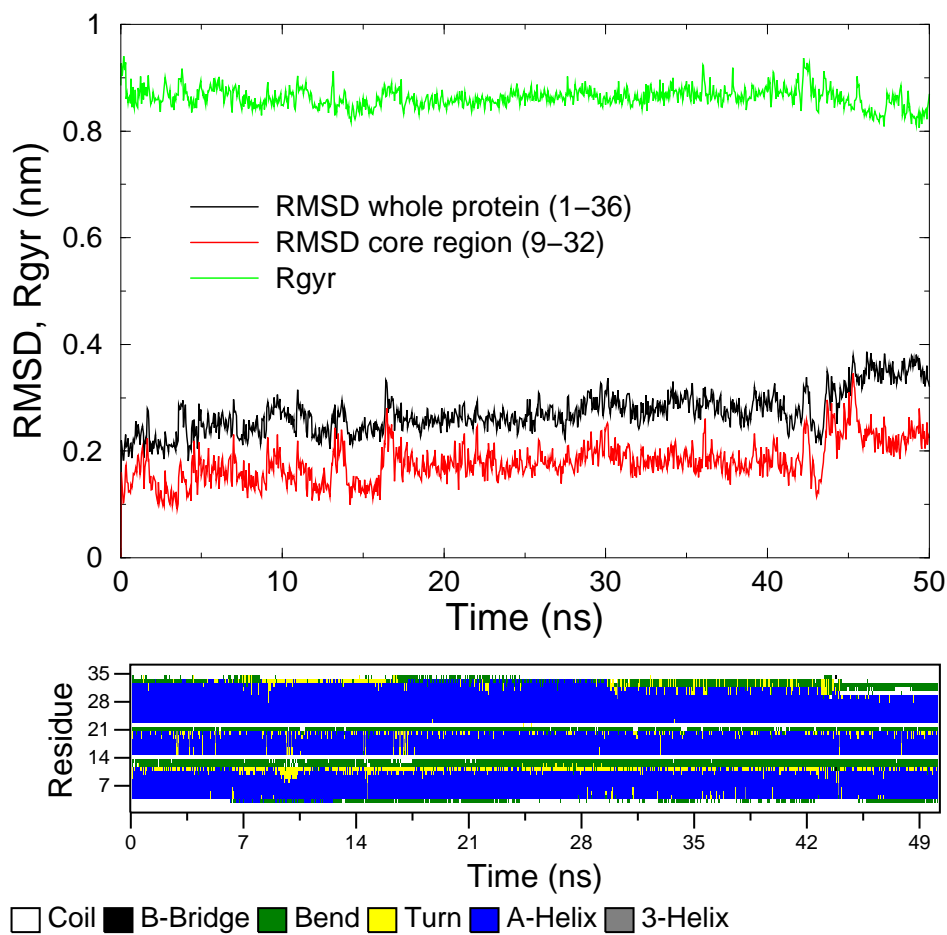}
\caption{All--atom MD structural characterization of HP36 native
conformation. Top: time evolution of the radius of gyration (green),
RMSD of the whole protein (black) and of the core region (red)
calculated against the average minimized NMR structure. The core
region comprises residues 9--32. Bottom: time evolution of the
secondary content (DSSP criterion).}
\label{native}
\end{figure}

Seven uncorrelated conformations were thus chosen from a MC
trajectory thermalised at $T_c$ and subject to the fine--graining
reconstruction procedure described in Material and Methods. The
all--atom protein models obtained were first solvated using the SPC
water model and subjected to minimization in order to remove possible
bad steric contacts. After equilibration, as described in Materials
and Methods (section Molecular Dynamics) each model was evolved with
completely unbiased MD simulations at 300 $K$ based on the Gromos96
force field and explicit solvent. To test the viability of the force
field at this temperature, and also to produce a term of reference for
the other simulations we first followed the dynamical evolution of
HP36 for 50 ns starting from the minimized averaged NMR model of the
PDB structure 1VII. The evolution shows a marked stability of $R_g$
and of the overall protein structure. The average RMSD of the core
region (residues 9 to 32) was about 0.2 nm, while for the whole
protein was 0.3 nm, consistently with the high core stability already
pointed out by Duan and Kollman (see Figure ~\ref{native})
\cite{Duan98}. In accord with this study and more recent ones
~\cite{Shirts}, the dynamical evolution appears to entail a
substantial movement of the helix at the N terminal which is
responsible for the RMSD increase in the last 7 ns of the trajectory.

For reasons of brevity, of the seven simulations (F1, F2, ... F7)
carried out starting from the reconstructed MC conformations we shall
typically concentrate on four of them: F1, F3, F4 and F7, which
exemplify the variety of observed dynamical behavior.

\begin{figure}[h!]
\includegraphics[width=0.7\textwidth]{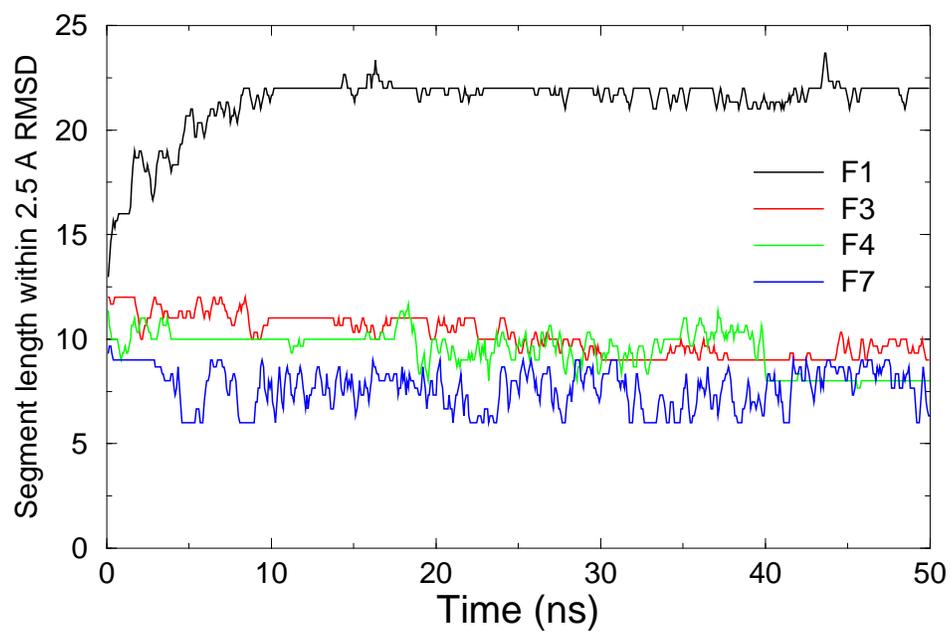}
\caption{Time-evolution, during simulations F1, F3, F4 and F7, of the
length of the largest protein fragment with an RMSD smaller than
2.5\AA\ from the native structure.}
\label{fig:window}
\end{figure}

To summarize and illustrate the system dynamics we report the time
evolution of several quantities including the radius of gyration, the
RMSD from the native state. Several clues about the dynamical behavior
are further obtained by inspecting the evolution of helical content in
the protein (measured by means of the DSSP criterion). In this
context, when referring to the native helical content, we shall follow
the notation of Duan and Kollman and denote with H1, H2 and H3 the
regions involving residues 4--8, 15--18 and 23--30, respectively. As a
further characterization of the degree of nativeness we have also
calculated, at various stages of the simulation, the length of the
largest protein segment(s) whose RMSD from the native state is lower
than 0.25 nm (see Fig. \ref{fig:window}). This cutoff distance was
chosen because it represents an upper limit for the structural
heterogeneity observed among conformations compatible with the
experimental indications. In fact, within this threshold nearly all
the NMR models are compatible with each other (while the maximum RMSD
deviation between the reference structure of the 1VII PDB entry and
any other NMR models is 2.1 \AA).

\begin{figure}
\includegraphics[width=0.7\textwidth]{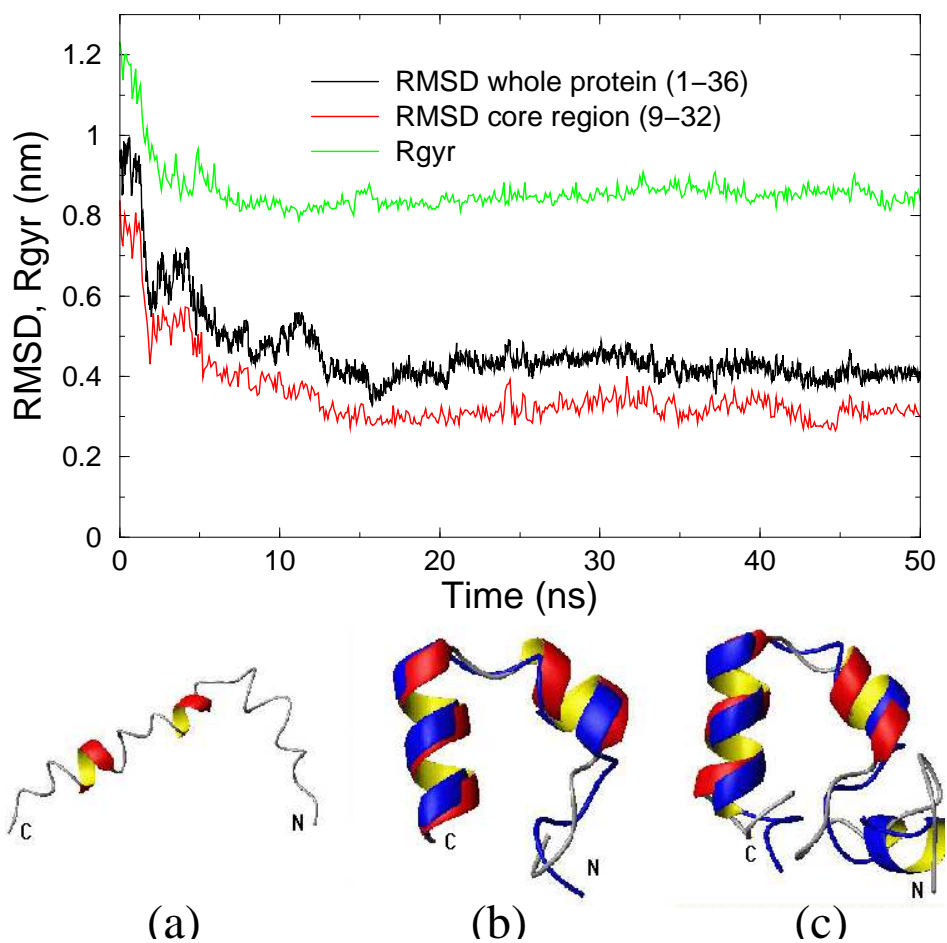}
\caption{Top panel: time evolution during simulation F1 of the radius
of gyration (green), of the RMSD over the whole protein (black) and
over the core region (red). Lower panel: (a) the starting structure of
simulation F1; best native structural alignment over the (b) core-region
and (c) entire protein of the representative structure of trajectory
F1. The representative is colored in red while the native reference
conformation is blue.}
\label{F1}
\end{figure}

\begin{figure}
\includegraphics[width=0.7\textwidth]{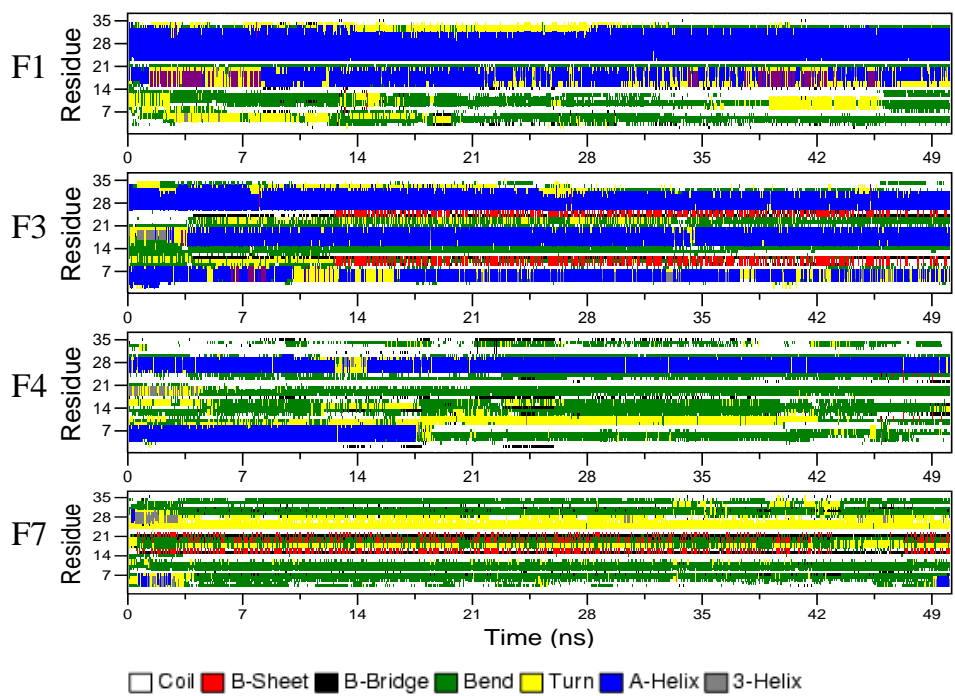}
\caption{Secondary structure time evolution for simulations F1, F3, F4
and F7. The DSSP criterion is used to define secondary structure
motifs.}
\label{dssp}
\end{figure}

As visible in Fig.  ~\ref{F1} the structure evolving in simulation F1,
after a short equilibration time, undergoes a rapid compaction as
shown by the decrease of $R_g$ from an initial value of 1.2 nm to
about the native one, $R_g^N \approx 0.89$ nm. The collapse in F1 is
not only rapid, but also ordered and productive of tertiary structures
yielding a core RMSD of 0.27 nm from the minimized average NMR
determined structure (Fig. ~\ref{F1} lower panel). The number of
consecutive residues whose RMSD is less than 0.25 nm from the NMR
determined structure is stable around 22 residues, that is about 60\%
of the whole structure (Fig. \ref{fig:window}). The helical content
corresponding to segments H2 and H3 is well conserved during the
simulation time (Fig. ~\ref{dssp}). The flexibility of the
N-terminal region prevents the formation of an optimal helical
geometry of the H1 segment. The RMSD value of the whole sequence with
respect to the NMR determined structure stabilizes at values around
0.40 nm.  The minimum RMSD value calculated over the whole ensemble of
29 model structures respecting the whole set of NMR restraints is
as low as 0.23 nm for the core region and 0.35 nm for the whole
protein.

\begin{figure}[h]
\includegraphics[width=0.7\textwidth]{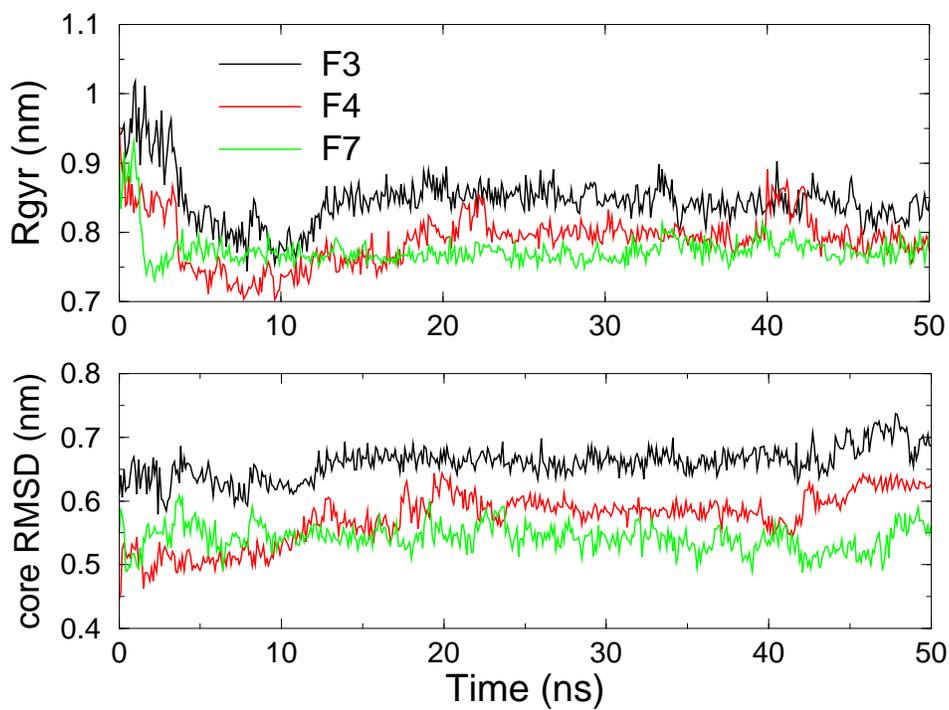}
\caption{Time evolution of radius of gyration (top) and core-RMSD
(bottom) of HP36. The RMSD was calculated against the average
minimized NMR structure of HP36.}
\label{F3F7}
\end{figure}

The starting structures of the other simulations were substantially
more compact than F1 with $R_g$ values of about 1 nm (see Fig
~\ref{F3F7}). The selected structures typically display a
non-negligible content of secondary structure, as visible in
Fig. ~\ref{dssp}. The inspection of overall native similarity
parameters, such as the RMSD, shows that a number of these runs does
not exhibit, within the simulated time span, any trend of systematic
approach of the native structure or native basin (Fig ~\ref{F3F7}).

Trajectory F3 displays a particularly interesting behavior. The DSSP
analysis reveals that helices H1, H2 and H3 are correctly formed
(Fig. ~\ref{dssp}). However, within the simulated time span the
trajectory is not approaching the native conformation since the RMSD
from the native state stabilizes around 0.6 nm (Fig ~\ref{F3F7}, black line). The secondary elements, while formed in the correct
native helical regions, assemble in a non-native geometry mainly due
to the formation of a stable hydrophobic core involving Phe11, Leu21,
Trp24 and Leu29.

A markedly different behavior is recorded in e.g. run F4 where (1) the
initial native similarity is acceptable, the core RMSD core being
$\approx 0.45$ nm (Fig ~\ref{F3F7}), (2) there is a good helical
content in H1 and H3 and (3) there are stretches of more than ten
consecutive residues below the RMSD threshold of 0.25 nm. This
initially promisingly similarity is, however, rapidly lost in a few ns
of dynamical evolution, eventually leading to negligible helical
content and overall native similarity. Interestingly, the loss of
native structure content is paralleled by the formation of a turn
conformation involving residues 8--10 and the paring of $\beta$--like
structures involving residues 2--7 and 10--15.

\begin{figure}[h!]
\includegraphics[width=0.7\textwidth]{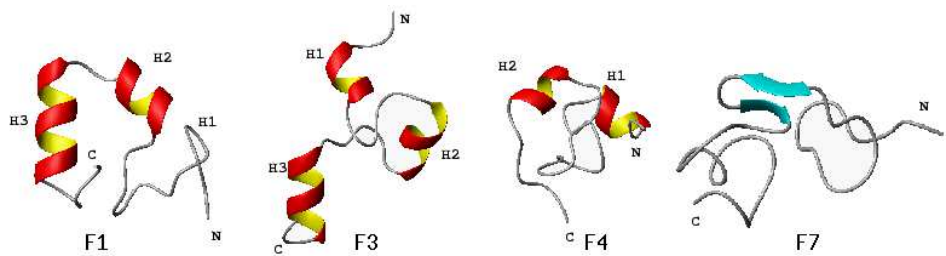}
\caption{Representative conformations of the most populated
conformational clusters in simulations F1, F3, F4 and F7.}
\label{representative}
\end{figure}

The tendency of the same segments to form $\beta$ structures is
further observed on trajectories starting from conformations without
significant helical content. This effect is here exemplified by run F7
(see Figs. \ref{F3F7}-\ref{representative}).

In summary, on the simulated time-scales, all--atom MD runs starting
from different selected structures display rather different behaviors
depending on the initial degree of compactness and native helical
content. In particular, the rapid progress towards the native villin
structure was observed in the case where the starting conformation was
not too compact and with good helical content in regions H2 and H3. On
the other hand, if the initial structure is rather compact and poor in
helical content then hydrophobic clusters are likely to form since
they lead to a rapid lowering of the protein internal
energy. Interestingly, the latter remains higher than the one achieved
in F1 or in the protein native state. In fact, the average internal
energy of runs F2--F7 is in the range [-1078,-1021] KJ/mol while for
run F1 and the native structure we observe -1109 and -1094 KJ/mol,
respectively. Moreover, the ``eager'' energy gain brought about by the
hydrophobic clusters leads to structures with several non-native
long-range contacts (in terms of the sequence separation). This fact
implies that a further progress towards the native state requires
major structural rearrangements.

\begin{figure}
\includegraphics[width=0.7\textwidth]{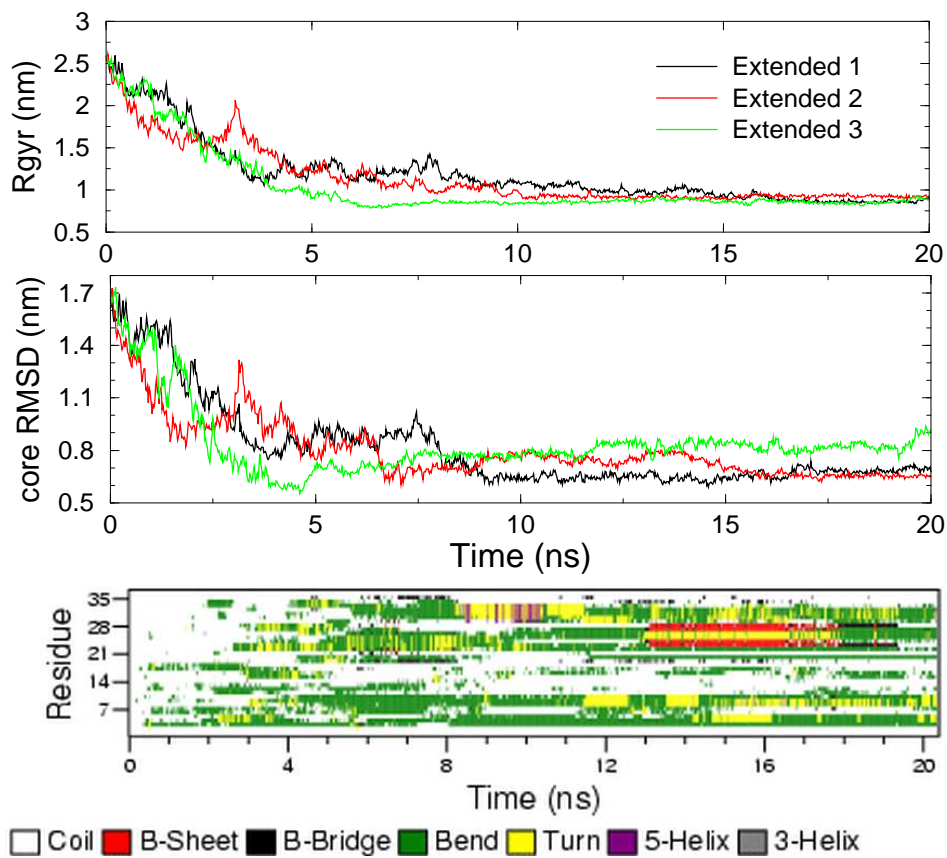}
\caption{Time evolution of radius of gyration (top) and core-RMSD
(middle) of HP36 in three different runs starting from the
fully-extended state. Bottom: typical time evolution of the secondary
content during simulations starting from the extended conformation.}
\label{extended}
\end{figure}

Besides the native trajectory, we have considered another reference
case constituted by runs started from fully extended conformation of
HP36 (Fig. ~\ref{extended}). These runs are important to check the
possibility to reach states with some degree of nativeness from
completely open conformations. In all the cases examined, the extended
structures rapidly collapse from the all--extended conformation to
compact ones in which no secondary or tertiary structure ordering
occurs on the same MD time scales of trajectories F1-F7. These runs
were stopped after 20 ns since MD at 300 K is unable to induce the
conformational changes necessary to break the compact and disordered
globular structures obtained, and because the number of required water
molecules made computation extremely slow. In all the simulations, the
structures become more compact because of solvent exclusion from the
contact with hydrophobic side--chains. This sort of hydrophobic
collapse, at least in the time spans analyzed here, is not enough to
bring about the ordering of local structures needed to move in the
proximity of the native basin.

\section*{Discussion} 

The present study aims at improving the reach of ordinary MD folding
simulations by recoursing to a preliminary, simplified, Monte Carlo
exploration of the free-energy landscape. If the free-energy landscape
associated to the simplified MC dynamics retained all the relevant
features of the true one, we could expect that the all-atom MD
simulations started from the sampled conformations are expected to
have significant advantages over, e.g. those starting from fully
extended protein configurations which can be affected by significant
lag-phases\cite{Fershtdistributed}. The advantage could, however, be
more conspicuous if the MC procedure allowed to pick structures from
the protein transition state ensemble, in which case the all-atom
dynamics would progress towards the native ensemble in a time scale
much shorter than the protein typical folding time.

Thus, the expectation is that the two-stage MC--MD trajectories are
``time-advanced'' with respect to those started, e.g. from extended or
other subjectively-chosen conformations. The quantification of the
expected time-advancement would be transparent and straightforward if
the folding process could be characterized in terms of one (or more)
reaction coordinates. All the natural and intuitive order-parameters
(such as $R_g$, RMSD, buried hydrophobic surface etc.) however appear
inadequate to this scope due to their wide degree of fluctuation
already observed by Duan and Kollman (and found also in the present
study) at all stages of their 1$\mu$s-long folding trajectory.
Therefore, the benefits of the present strategy can be ascertained
only through the comparison of the degree of order and ``nativeness''
of the explored trajectories compared to that achievable by
e.g. starting from extended conformations.

Therefore, we first discuss the reference cases constituted by the
runs carried out starting from the native structure and from extended
conformations. Over the simulated time scale, the native run appears
to be very stable, the only appreciable structural changes involving
the mobile regions outside the protein core. We have monitored the
stability of the local and non-local native contacts. Two amino acids
are said to be in contact if the spatial separation of their
$C_\alpha$'s is below 7.5 \AA. We classify as local contacts those
involving residues at a sequence separation equal to three or four
(i.e. we omit from the count the non-informative contribution of
separation 1 or 2).  The native structure possesses 52 distinct local
contacts, while the non-local ones, pertaining to sequence separation
greater than 5 are 13. In the native run the fraction of local native
contacts maintained during the simulation has the nearly constant
value of $q_{l} = 0.9$. The non-local one, $q_{nl}$ instead, due to
the mobility of the N-terminus, oscillates between 0.55 and 0.75. The
overall RMSD from the native structure remains low and compatible with
what found in ref. \onlinecite{Duan98}, despite the use of a different
force-field.

The high RMSD observed in runs started from the extended structures
are obviously reflected in low values for both $q_l$ and $q_{nl}$. In
particular, $q_l$ typically fluctuates between 0.2 and 0.3, while
$q_{nl}$, is practically zero. This does not imply that long-range
contacts are absent. On the contrary, as mentioned in the results
section, all extended runs evolve rapidly towards globular
conformations, and hence with non-local contacts, with formed
hydrophobic clusters (different across the runs).\\ The coarse-grained
structures sampled at the collapse temperature, $T_c$, have various
degree of compactness with $R_g$ ranging between 8 and 13 \AA.  As a
consequence the sampled conformations have a good degree of structural
heterogeneity, since the average RMSD between any pair of these
structures is 5.9 $\pm$ 1.2 \AA, while the typical RMSD against the
core of the villin headpiece is 7.1 $\pm$ 1 \AA. Out of these ensemble
ten structures were randomly selected; since their mutual RMSD was not
inferior to 6 \AA they represent well the structural diversity
encountered at $T_c$. After the reconstruction and energy-minimization
procedure these structures were used as inputs for the MD runs. All
these starting conformations had a degree of local native similarity
substantially higher than what achieved in the extended runs, as
summarized in table \ref{mdtable}. The observed degree of native
similarity is not the result of a ``blind'', featureless,
compactification, but reflects the fact that the coarse-grained model
indeed captures some of the physico-chemical forces that govern the
folding process. To ascertain this we have applied the same MC scheme
on a sequence obtained by a random reshuffling of the villin
one. Despite the preservation of the native composition, which must
reflect in an overall bias towards local helical conformations, the
native similarity at $T_c$ in this second situation is appreciably
different. In fact, the average values of $q_l$ and $q_{nl}$ are 0.25
and 0.08, respectively, while the average RMSD against the protein
core is 8.2 \AA.

On the contrary, $q_l$ ranged from 0.3, for F6, to more than 0.6, for
cases F1 and F3 which had partially-formed helices in the correct
regions. In these two runs, the initially good local similarity is
further improved in about 10 ns of evolution, reaching $q_l \approx
0.8$. As pointed out in the results section, despite the presence of
well-formed helices the tertiary organization found in the two runs is
distinctly different. This is reflected by $q_{nl}$ which starts from
a negligible value in both cases and reaches again in about 10 ns of
evolution the excellent peak value of 0.75 for F1 (and then
stabilizing to 0.7), while it remains virtually zero for the whole F3
run. The case of two trajectories containing structures with high
native local content but markedly different non-local one provides an
appealing illustration of the role played by the interplay of local
and non-local interactions for steering the folding process. In fact,
the establishment of the correctly-formed non-local contacts appears
to be crucial for providing the necessary stability to maintain both
the native secondary and tertiary organization. In fact, after the
first 10 ns of dynamical evolution, both $q_l$ and $q_{nl}$ settle to
the stable values of 0.7 and 0.45, respectively. On the contrary, due
to the lack of a native (correct) non-local scaffolding the secondary
content in simulation F3 is steadily degraded, dropping from 0.8 to
0.5, in the subsequent evolution. A similar erosion of the native
local similarity is further encountered for simulation F4. The
starting value of $q_l \approx 0.5$ which falls to values of about
0.25 due to the lack of a substantial native non-local network of
interaction (the typical value of $q_{nl}$ being about 0.3). The
evolution of trajectory F7 is, among all those started from MC-sampled
configurations, the one that has the worst performance, since the
average value of $q_l$ is only about 0.3 while $q_{nl}$ is negligible.

The analysis of several dynamical parameters and the inspection of
structures sampled from the MD runs shows that there are several
mechanisms responsible for thwarting the dynamics towards states with
low native similarity. For trajectory F3, the inability to attain the
correct helical packing is due to the formation of a hydrophobic
cluster involving Phe11, Leu21, Trp24 and Leu29 which traps the system
and prevents its further evolution on the simulated time-scale.  The
cases of the other trajectories that, on the simulated time scale, do
not evolve towards the native basin is somewhat different from the
previous one. It appears that a systematic mechanism that opposes and
destabilizes the formation or packing of the native helices in the
native fold is the tendency to form $\beta$-sheets. In fact,
concomitant with the disruption of helical content (especially of the
``frail'' H1 helical region) we have observed the systematic
appearance of extended regions (strands) in the protein. This is
visible, for example, in Fig. ~\ref{dssp} and ~\ref{representative}
where the existence of extended conformations as well as their
stability under subsequent dynamical evolution is evident. A more
precise insight in the mechanism leading to $\beta$-sheets is obtained
by examining the sequences of the residues taking part into the
strands. Both sequence 2--7 and 10--15 contain hydrophobic residues
such as Phe7, Phe11, Met 13 and Val10. These residues have unfavorable
interactions with water, and one mechanism to minimize the contact is
the packing of hydrophobic side--chains in adjacent extended regions
of $\beta$-structure.

Moreover, the short stretch of the three native helical segments, is
such that each individual helix is not expected, on the basis of
helix-coil transition models, to be stable on its own but ought to
rely on additional mid- and long-range interactions in terms of
sequence separation. The importance of optimal mid- and long-range has
been recently verified experimentally by Raleigh and coworkers
~\cite{Raleighpeptidevillin}. These authors systematically studied
through CD and NMR spectroscopy peptides spanning different regions of
HP36, and showed that only when tertiary interactions are present in a
sequence comprising at least two helices, a significant amount of
ordered structure is present.

In this picture, a generic hydrophobic collapse is thus not enough to
determine the onset of a sufficient number of native--like
interactions necessary for driving the system in proximity of the
native basin. A different kind of structural organization appears to
be required, based on the harmonious interplay of non-local
interactions with the local ones favoring helical conformations
\cite{anfinsenscheraga}. This picture is consistent with the topomer
search mechanism which assumes that the rapid progress towards the
native state is possible only after the establishment, in a diffusive
conformational search, of a minimal set of contacts embodying the main
traits of a protein native topology \cite{topomer99}. In this scheme a
paramount role is played by mid-range interactions
\cite{anfinsenscheraga,topomerPlaxco,topomer99,Shortle01}.

The emphasis here is that, for the case of the villin headpiece, the
structures leading to this correct interplay of short and long-range
contacts can be generated through a preliminary stochastic search of
the simplified free-energy landscape. All-atom and explicit-solvent MD
runs starting from these configurations present significant advantages
over the traditional choice of extended conformations. In fact, the
regions in structure-space visited by the seven trajectories
considered here are on average much closer to the native basin than
for extended runs. Not only there is a benefit in terms of the quality
and variety of the trajectories, but there is a further advantage in
terms of computational expenditure. In fact, the necessity to take
explicitly into account the solvent during the simulation
~\cite{Wolynesacqua}, implies that a very large number of water
molecules need to be considered in the simulation cell enclosing
extended starting simulations. Though this number can be decreased as
the protein compactifies, it impacts severely on the total simulation
time so that reaching the same-simulation time of e.g. 20 ns each
extended run typically requires a sixfold CPU increase over cases
F1-F7.

The validity and power of the method discuss here is revealed not only
by the typical improvement of the two-stage runs against the extended
case, but especially by the ability to reach and maintain the villin
HP36 core during simulation F1. This success reflects the viability of
both the starting structure picked by the MC search and of the MD
force field. In fact, as visible in Fig. ~\ref{F1} the F1 starting
structure is not similar to the native state but it presents the
possibility of rapidly evolving towards it due to the presence of two
partial native helices in an overall non-compact conformation. It is
the action of the force-field that makes this progression possible (of
course after the reconstruction of the atom detail from the coarse
model).

The representative structure of run F1 (determined through the application of the structural clustering method described by Daura and coworkers~\cite{Daura99a}) has an RMSD over the core
region of 0.27 nm from the minimized average NMR structure, and a
minimum of only 0.24 nm over the whole set of 29 NMR models. This
represents an improvement over the landmark result of Duan and Kollman
\cite{Duan98} who obtained an RMSD of 0.4 nm with a threefold increase
of simulation time. The significance of the degree of nativeness
obtained here for the HP36 core is highlighted by two facts. First,
the core RMSD is smaller that typical thresholds used to define a
successfully-folded conformation in similar contexts (and for
analogous lengths)\cite{Snow02}. Secondly, it should be borne in mind
that the 29 NMR models of the Villin headpiece have a non-negligible
degree of structural heterogeneity among themselves (up to nearly 3
\AA), with the maximum RMSD difference between the minimized average
NMR structure (PDB code 1vii) and any other model being 0.21 nm.

To summarize, we have followed the all-atom dynamical evolution of the
Villin headpiece starting from several configurations obtained by a
coarse exploration of the system energy landscape. The seven simulated
trajectories included a notable instance where a correctly-folded
configuration of the protein core was reached and
maintained. Furthermore, the significant secondary content and
organization found in all starting structures resulted in interesting
dynamical evolutions that, even when not progressing towards the
folded state, convey valuable information on the folding process, as
the trapping mechanism associated with the formation of contacting
strands. The protocol used here has been kept as general and unbiased
as possible and may be further extended e.g. at the level of the
coarse-grained model and/or of the selection criteria of the starting
structures. Therefore, the proposed strategy ought to be applicable to
other instances of short proteins with significant advantages over the
use of extended structures as the default unbiased starting point of
all-atom dynamical simulations with explicit solvent.

\section*{Materials and methods}

\subsection*{Molecular Dynamics simulations and analysis}

{Every all--atom Molecular Dynamics Experiment was run using the same
simulation protocol and conditions.  In the simulations starting from
the NMR, completely extended and reconstructed conformations derived
from the preliminary MC analysis (runs F1 to F7), all basic NH\(
_{2}\) groups were considered protonated while all the acidic COO\(
^{-} \) groups were considered deprotonated. These conditions resulted
in a total charge of \( +2\) on the protein. The protein in each
simulation was solvated with explicit water molecules in a period
octahedral box large enough to contain the protein and \( 1.0 \) nm of
solvent on all sides. The simple point charge (SPC)~\cite{Berendsen87}
water model was used to simulate the solvent. Electroneutrality of the
system was ensured by the addition of two negatively charged chloride
ions. Table ~\ref{mdtable} reports the number of water molecules in
each simulation.  The same equilibration protocol was used in all
simulation:\\
\begin{description}
\item[1.]{Each system was initially energy minimized with a steepest descent
method for \( 1000 \) steps. The calculation of electrostatic forces
utilized the PME implementation of the Ewald summation method ~\cite{pme}.  The
GROMOS96 force field~\cite{vanGunsteren98,GROMOS96} was used.}
\item[2.]{To release excess strain in simulations F1--F7 due to possible artifacts in the reconstruction procedure, three sequential MD runs of \( 50 \)~ps each with position restraints on the protein, backbone and sidechains, respectively. In the first, the protein atoms were kept fixed with the solvent molecules free to move. In the subsequent run, the backbone of the protein was kept fixed with both the sidechains and solvent free to move in MD. In the third and final simulation, the sidechains were kept fixed, while the backbone and solvent atoms were allowed to move.}
\item[3.]{After these first relaxation steps, each system was heated to 300K by \( 200 \)~ps of MD simulation under NPT conditions, with no restraints, by weak coupling to a bath of                   
constant pressure (P\( _{0}=1 \) bar, coupling time \( \tau _{P}=0.5 \) ps)~\cite{Berendsen84}.}
\item[4.]{Each of the systems was then equilibrated by \( 50 \) ps of unrestrained MD with coupling to an external temperature bath~\cite{Berendsen84} with coupling constant of \( 0.1 \)~ps.}
\item[5.]{The production runs at 300K, using NPT conditions were \( 50\) ns long in the case of the simulation starting from the NMR determined structure 1VII.pdb and simulations F1 to F7. Due to the extremely high number of explicit water molecules, the
three independent MD simulations starting from the totally extended
conformation were run for 20 ns. In all production runs, the temperature was maintained close to the
intended values, by weak coupling to an external temperature bath
~\cite{Berendsen84} with a coupling constant of \( 0.1 \)~ps. The
protein and the rest of the system were coupled separately to the
temperature bath.  The LINCS algorithm ~\cite{Hess97} was used to
constrain all bond lengths.  For the water molecules the SETTLE
algorithm ~\cite{Miyamoto92} was used. A dielectric permittivity, \(
\epsilon =1 \), and a time step of \( 2 \)~fs were used.
All atoms were given an initial velocity obtained from a Maxwellian
distribution at the desired initial temperature of 300K.}
\end{description}}

The extended conformation of the protein was generated with
Swiss-PdBViewer ~\cite{SVIZZERO} and the system was solvated with the
same protocol used for the folding simulations.

All the minimization, MD runs and analysis of the trajectories were
performed using the GROMACS software package.~\cite{GROMACS} The
graphical representations of the protein were realized with the
program MOLMOL~\cite{Koradi96}.

\begin{table}[htbp]
\begin{tabular}{ c c c c c }\hline
Simulation & Water Molecules & $q^0_{l}$ & $q^0_{nl}$ & initial core RMSD\\  \hline\hline
F1 & 6568   & 0.67 & 0.1  & 8.4\\
F2 & 5187   & 0.50 & 0.05 & 8.5\\
F3 & 3892   & 0.61 & 0.15 & 6.1\\
F4 & 3140   & 0.50 & 0.10 & 4.5\\
F5 & 2466   & 0.42 & 0.23 & 4.8\\
F6 & 2848   & 0.31 & 0.11 & 7.0\\
F7 & 2911   & 0.46 & 0.04 & 5.5\\\hline
E1 & 11661  \\
E2 & 8560 \\
E3 & 7272 \\\hline
Native & 2768 \\\hline \hline
\end{tabular}
\caption{Summary of the main feature of the various MD runs Number of
solvent particles in each of the simulations. For trajectories F1-F7
are also provided the fraction of local and non-local native contacts
at the beginning of the MD run, as well as the initial RMSD over the
core of the protein (residues 9-32).}
\label{mdtable}
\end{table}

\subsection*{Monte Carlo}

The starting configurations for all-atom MD simulation were obtained
from a preliminary Monte Carlo exploration of the configurational
space. In order to make this preliminary search as efficient as
possible we have adopted a simplified representation of protein
conformations. Most of the procedures adopted to coarse grain the
microscopic degrees of freedom of proteins substitute a whole amino
acid with a small number of effective centroids. We have followed this
choice and used effective $C_\alpha$ and $C_\beta$ centroids for each
amino acid (with the exception of Gly which lacks the $C_\beta$
centroid). This choice brings about a drastic simplification for the
structure of the Hamiltonian which will involve only interactions
between the centroids. In particular we regard the $C_\alpha$
centroids as the centers of the effective interactions among
contacting residues. The $C_\beta$ ones play, instead, a passive role
being used to capture the steric hindrance of the sidechains. In order
to minimize the dimensionality of conformational space only the
$C_\alpha$ degrees of freedom where considered and the effective
$C_\beta$'s were constructed according to the Park and Levitt rule
\cite{cb_construct}.

The dynamics in conformation space is carried out using a Monte Carlo
technique. At each attempted Monte Carlo step the current conformation
($C_\alpha$-trace) is distorted by means of crankshaft and pivot
rotations. These types of moves are employed since they do not alter
the separation of consecutive $C_\alpha$'s (bond length).  The newly
generated conformation is then accepted/rejected according to the
standard Metropolis rule. The system energy function comprises the
following terms:

\begin{equation}
{\cal H} = {\cal H}_{steric} + {\cal H}_{pairwise}+ 
{\cal  H}_{cooperative} +  {\cal H}_{chiral} \ . 
\end{equation}

\noindent The first term, ${\cal H}_{steric}$, is used to enforce some
basic steric constraints on the conformations generated with the MC
procedure. These knowledge-based constraints involve both two- and
three-body terms. In particular, no two centroids of distinct
residues, $i$ and $j$ (be they $C_\alpha$ or $C_\beta$) are allowed to
come at a distance smaller than 3.0 \AA, with the proviso that if $|
i-j|=2$ then their minimal separation is set to 5 \AA.  These pairwise
constraints needs to be complemented by suitable three-body terms for
a more accurate treatment of steric hindrance. In fact, it has been
shown that proteins are well-described as ``tubes'' with a uniform
effective thickness of about $\Delta=2.7$ \AA \cite{opthelix,secstr}.
This implies that the local radius of curvature of the backbone (the
$C_\alpha$ trace) cannot be smaller than $\Delta$ and that two
distinct backbone portions cannot come at a separation smaller than
$2\Delta$. From a numerical point of view these requirements are
enforced by ensuring that the radius of the circumference going
through any triplet of distinct $C_\alpha$'s is greater than $\Delta$.

The second energy term, ${\cal H}_{pairwise}$ is used, instead, to
capture the effective interactions among pairs of sufficiently closed
residues.  The strength of the contact interaction between two
residues $i$ and $j$ is modulated with a sigmoidal function,
$f(r_{i,j})$, which depends on the separation of the corresponding
$C_\alpha$ centroids, $r_{i,j}$, and has a a point of inflection at
6.5 \AA,

\begin{equation}
f(r_{i,j})= {1 \over 2} + {1 \over 2} \tanh(6.5 - r_{i,j})
\end{equation}

\noindent Therefore, ${\cal H}_{pairwise}$  can be written as

\begin{equation}
{\cal H}_{pairwise} = \sum^\prime_{i,j} \epsilon(S_i,S_j) \, f(r_{ij})
\end{equation}

\noindent where the prime denotes that the summation runs over
non-consecutive residues, $S_i$ denotes the type of residue at the
sequence position $i$ and $\epsilon$ is the symmetric matrix
describing the strength of the effective residue-residue interactions.
Various criteria have been employed to extract tables of effective
potentials. Here we recoursed to the one developed by Kolinsky, Godzik
and Skolnick \cite{KGS} which proved useful to identify the correct
fold of short proteins \cite{Kolinskyskolnick94a}. In addition, in
analogy with ref. \onlinecite{Kolinskyskolnick94a}, we have introduced a
cooperative four-body interaction meant to favor the packing of
secondary motifs:

\begin{eqnarray}
&{\cal H}_{cooperative} =\\ \nonumber
& \sum_{i,j,a,b} f(r_{i,j}) \, f(r_{i+a,j+b})
[\epsilon(S_i,S_j)+\epsilon(S_{i+a},S_{j+b})]/20
\end{eqnarray}

\noindent where $a$ and $b$ can take on the values $\{\pm 3, \pm 4\}$.
This term introduces an additional energy contribution if both bonds
$(i,j)$ and $(i+a,j+b)$ are present. 

\begin{figure}
\includegraphics[width=0.5\textwidth]{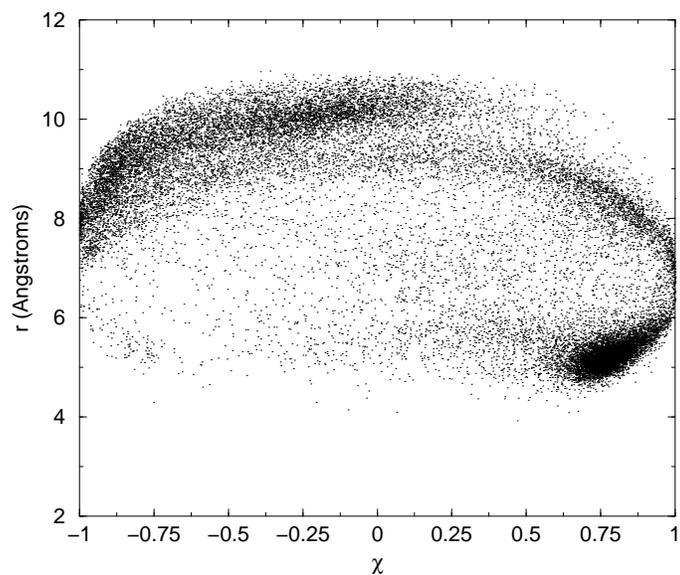}
\caption{Scatter plot of the chirality and end-to-end distance
  of fragments taken from the representative structures of PDBselect.}
\label{fig:chirality}
\end{figure}

Besides these interactions we have further found useful to reduce the
structural freedom of the coarse-grained conformations by introducing
a knowledge-based chiral potential, ${\cal H}_{chiral}$. The chiral
bias and constraints were derived by parsing several structurally
independent proteins into fragments of 4 consecutive residues. For
each fragment we computed the end-to-end distance, $r$, and the
associated chirality $\chi = (\vec{b}_{12} x \vec{b}_{23}) \cdot
\vec{b}_{34}$ where $\vec{b}_{ij}$ denotes the normalized vector
joining the $i$th and $j$th $C_\alpha$ centroids in the fragment. The
scatter plot $\chi$ vs $r$ of Fig. \ref{fig:chirality} shows that some
regions (in principle compatible with feasible configurations) are
forbidden, or very rarely populated. This fact is used to preclude
these regions from being explored by the MC dynamics. More precisely,
any fragment ($i$,$i+4$) of any configuration encountered in the
stochastic dynamics has to avoid the following regions of the $\chi,r$
plane (the constraints are enforced again through a large energy
penalty): (a) $\chi <0$ and $r < 7.5$\AA, (b) $r > 11$ \AA, (c) $r <
4$\AA.  It is also possible to notice that a large number of
configurations cluster around the point ($\chi \approx 0.7, r \approx
5.5$\AA) which corresponds to helical conformations; the other highly
populated band corresponding to $r > 9$\AA is instead associated
with extended conformations.

In the absence of any chiral bias, besides the constraints mentioned
above the MC sampling tends not to occupy the helical region even for
sequences with strong helical propensities. We therefore introduced a
sequence-dependent bias to favor the occupation of the helical region
in such cases. This is accomplished by processing the set of proteins
representatives (obtained from the PDBselect list
\cite{pdbselect1,pdbselect2}) to calculate the probability of each
amino acid to be in fragments having helical (H) or extended (E)
conformations in the various protein instances. The calculated values
are given in Table \ref{tab:propensity} (notice that since each amino
acid can occupy other portions of the phase diagram, $p_H$ and $p_E$
do not sum to 1). Within the MC scheme, this knowledge-based
information is exploited by calculating, for each fragment of four
consecutive residues, the value of helical and extended probability,
$\bar{p}_H$ and $\bar{p}_E$, averaged over the four residues. The
occupation of the helical region of the $\chi-r$ plane is favored if
$\bar{p}_H$ is sufficiently high and, concomitantly, $\bar{p}_E$ is
sufficiently low.

\begin{table}[htbp]
\begin{tabular}{| l | l | l || l | l | l |  } \hline
a.a.  &  $p_H$  & $p_E$  &  a.a. & $p_H$ & $p_E$ \\ \hline \hline
ALA & 0.45 & 0.19 & LEU &  0.40 & 0.24\\             
ARG & 0.38 & 0.21 & LYS &  0.35 & 0.20\\             
ASN & 0.28 & 0.18 & MET &  0.42 & 0.24\\             
ASP & 0.32 & 0.17 & PHE &  0.31 & 0.32\\             
CYS & 0.27 & 0.23 & PRO &  0.16 & 0.19\\             
GLN & 0.43 & 0.19 & SER &  0.30 & 0.22\\             
GLU & 0.43 & 0.18 & THR &  0.28 & 0.27\\             
GLY & 0.18 & 0.21 & TRP &  0.31 & 0.28\\             
HIS & 0.27 & 0.25 & TYR &  0.28 & 0.32\\             
ILE & 0.32 & 0.34 & VAL &  0.26 & 0.39\\\hline \hline
\end{tabular}
\vskip 0.5cm
\caption{Probability of the different amino acids to be in helical or
extended conformations. The probabilities, indicated with $p_H$ and
$p_E$, were calculated by processing a set of nearly 140 structures
determined by high-resolution Xray crystallography and reported in the
PDBselect list of protein representatives.}
\label{tab:propensity}
\end{table}

A test on an independent set of proteins has shown that a useful
criterion to identify putative helical fragments is to require that
$p_H (1 - p_E) > 0.25$ and $p_E(1 - p_H) < 0.15$. These inequalities
ensure that about 50 \% of helical fragments are correctly recognized,
while the probability that an extended fragment is mistakenly labeled
as helical is about 10\%. Each putative helical fragment is biased
towards occupying the helical region of the $\chi,r$ plane through the
following potential term:

\begin{equation}
V(\chi,r)={1 \over 2} + {1 \over 2} \tanh[{\chi - 0.7 \over \sigma_\chi}]\, \tanh [{5.5 - r \over \sigma_r}]\, .
\end{equation}

\noindent The values of $\sigma_\chi$ and $\sigma_r$ are set to 0.1
and 0.3, respectively to reflect the spread observed in the $\chi-r$
distribution of helical fragments. These potential biases are
accumulated in the energy term ${\cal H}_{chiral}$:

\begin{equation}
{\cal H}_{chiral} = V_0 \sum_l \, V(\chi_l,r_l)
\end{equation}

\noindent where the sum runs over all putative helical fragments of
four residues and the amplitude coefficient, $V_0$, is equal to 3.0 .
This value was chosen after applying the MC sampling strategy to
various short proteins (from about 30 to 50 residues) belonging to
different structural families, (PDB codes 1aie, 1ajj, 1ppt, 1ptq,
2erl). We stress again that, in our scheme, we have not found the
necessity to add any bias towards extended conformations for local
sequences that were not putatively helical. Indeed, for $\beta$
proteins of length comparable to the villin headpiece (e.g. 1eit,
1pmc) virtually no residues can be found in helical conformations at
all MC temperatures. Finally, the MC exploration was performed
starting from extended conformations at high $T$ and then gradually
lowering the temperature as in traditional simulated annealing
schemes. For $V_0 \approx 3.0$ the formation of secondary structures
occurred at about the collapse temperature (i.e. the temperature where
a rapid decrease of the radius of gyration occurs). In this situation,
the presence of secondary structures results from a
physically-appealing similar interplay of the sequence-dependent
chiral bias and of the more complicated energy terms.\\
In essence, the present Hamiltonian embodies the main ingredients that
are regarded as crucial in coarse-grained approaches to protein
folding, namely pairwise interactions, local biases for backbone
chirality and cooperative interactions to promote the packing of
secondary motifs. Therefore, the interesting and rich MD evolution of
the reconstructed coarse-grained structures is indicative that the
two-stage scheme can be profitably used under quite general
conditions, and not the necessity to use the present coarse-grained
energy functional. In particular a foreseable improvement to the
present Hamiltonian would be to substitute the cooperative term by a
suitable modeling of main-chain hydrogen bonds. This has, in fact,
proved to be an important ingredient to aid the formation of
beta-sheets structures in globular proteins \cite{kolsko04}.

\subsection*{Reconstruction procedure}

As explained above, all the protein conformations generated through
the Monte Carlo procedure have a simplified structural representation
essentially described by the $C_\alpha$ degrees of freedom. These
coarse-grained structures need to be reconstructed with all the atomic
detail before they can be processed in ordinary all-atom molecular
dynamics schemes.

The problem of reconstructing reliably a protein's atomic detail
starting from a coarse-grained representation has been addressed
previously. Several studies have appeared which address the problem of
reconstructing, with full atomic detail, a protein's backbone and/or
sidechains starting from very limited structural informations
\cite{reconstr_correa,reconstr_rey,holm_sander_reconstr,cohen_reconstruction,reconstr_scheraga}.
In the present study we have developed a novel knowledge-based
strategy that is simple and whose accuracy is not inferior to more
sophisticated techniques. We now illustrate this reconstruction
scheme. For reasons of space our description will be complete, but
schematic; a more detailed presentation of the method and its
performance will be published in the future.

The algorithm that reconstructs the full atomic detail of a given $C_\alpha$
trace is based on the use of a library of protein fragments built from
about 100 NMR gapless structures taken from the PDBselect
list\cite{pdbselect1,pdbselect2}. These structures were parsed into
template fragments of four consecutive residues retaining the whole
atomic detail of the mainchain and of the sidechains of the middle
residues. For each set of four consecutive $C_\alpha$'s in the $C_\alpha$ trace one
finds the best superimposable fragment in the library and assigns the
central peptide plane to the protein to be reconstructed after an
optimal roto-translation \cite{kabsch}; the first and last peptide
planes are treated separately. Finally, the sidechain of any given
residue, $R$, is obtained by first considering only the set of
template fragments where one of the middle residues, $R^\prime$, is of
the same type as $R$. Next, after aligning (in sequence) $R$ and
$R^\prime$, the sidechain of $R$ is assigned (again after an optimal
roto-translation) from the fragment providing the best superposition
with the reconstructed backbone.

\begin{figure}
\includegraphics[width=0.5\textwidth]{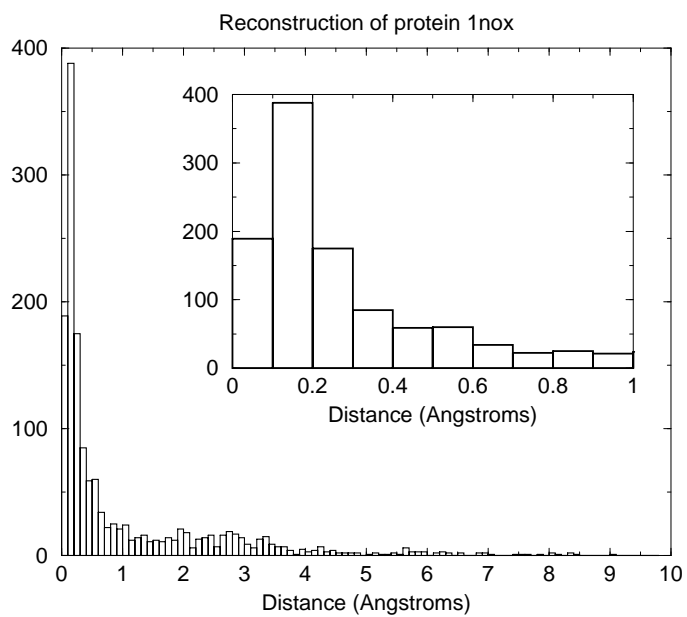}
\caption{Histogram for the distance of the reconstructed heavy-atoms
  from the crystallographic positions in protein 1nox.}
\label{fig:reconstruction}
\end{figure}

In Fig. \ref{fig:reconstruction} we have reported the distribution for
the errors in the reconstruction of known protein structures. It
appears that 70 \% of the heavy atoms are placed within 1 \AA\ of the
crystallographic positions (and 62 \% within only 0.5 \AA). Although
the typical performance is very satisfactory, it is worth pointing out
that, as for other methods, a wrong placement of ramified sidechains
can occasionally give rise to large deviations from the correct
positions (see the tail of the error distribution in
Fig. \ref{fig:reconstruction}). Large deviations may also be
encountered in correspondence of charged residues, whose sidechain
orientation is heavily influenced by the local electrostatic
(obviously not captured in our simple reconstruction scheme). This is,
however, a minor problem in this context, since the reconstructed
structure is solvated, energy-minimized and equilibrated before
starting the MD run.

\vskip 0.3cm 
{\bf Acknowledgments.} We acknowledge support from CNR, INFM and MIUR
Cofin 2003.

\newcommand{\noopsort}[1]{} \newcommand{\printfirst}[2]{#1}
  \newcommand{\singleletter}[1]{#1} \newcommand{\switchargs}[2]{#2#1}

\newpage

\end{document}